# Mild sonochemical exfoliation of bromine-intercalated graphite: a new route towards graphene

E. Widenkvist[1], D. W. Boukhvalov[2], S. Rubino[3], S. Akhtar[3], J. Lu[1], R. A. Quinlan[4], M. I. Katsnelson[2], K. Leifer[3], H. Grennberg[5] and U. Jansson[1]

[1] *Department of Materials Chemistry, Uppsala University, BOX 538, SE-752 21 Uppsala, Sweden*

[2] *Institute for Molecules and Materials, Radboud University of Nijmegen, 6525 ED Nijmegen, The Netherlands*

[3] *Department of Engineering Sciences, Division for Electron Microscopy and Nanoengineering, Uppsala University BOX 534, SE-752 21 Uppsala, Sweden*

[4] *Department of Applied Science, The College of William and Mary, 325 McGlothin Street Hall, Williamsburg, VA 23187, USA*

[5] *Department of Biochemistry and Organic Chemistry Uppsala University, BOX 576, SE- 751 23 Uppsala, Sweden*

E-mail: ulf.jansson@mkem.uu.se

## Abstract

A method to produce suspensions of graphene sheets by combining solution-based bromine intercalation and mild sonochemical exfoliation is presented. Ultrasonic treatment of graphite in water leads to the formation of suspensions of graphite flakes. The delamination is dramatically improved by intercalation of bromine into the graphite before sonication. The bromine intercalation was verified by Raman spectroscopy as well as by x-ray photoelectron spectroscopy (XPS), and density functional theory (DFT) calculations show an almost ten times lower interlayer binding energy after introducing $Br_2$ into the graphite. Analysis of the suspended material by transmission and scanning electron microscopy (TEM and SEM) revealed a significant content of few-layer graphene with sizes up to 30 μm, corresponding to the grain size of the starting material.

# 1. Introduction

Graphene is a two-dimensional form of graphite and consists of a single layer of carbon atoms in a honeycomb crystal lattice. The recent efforts to synthesize graphene have sparked much interest, due to the remarkable electronic properties of this material [1–3]. The primary method of graphene production is micromechanical cleavage of graphite [2, 3]. This method is difficult to control and also to scale up for industrial applications. Graphene is also synthesized by epitaxial growth on silicon carbide but this method requires high temperatures and the graphene can be difficult to transfer from the silicon carbide substrate [4]. An alternative chemical route is oxidation of graphite. However, the resulting graphene oxide is difficult to reduce completely to graphene [5, 6]. Recently presented chemical methods also include the use of surfactants or reduction to dissolve graphite [7, 8]. Here we demonstrate an alternative method combining intercalation and sonochemistry to fabricate graphene. This is a liquid-based method with the potential of being low-cost and readily scalable. Moreover, a solution-based method will allow for easy future manipulation by techniques used in

# 2. Results and discussion

## 2.1. Bromine intercalation

Bromine intercalates into graphite as molecular $Br_2$ [10], and to investigate the effect of the molecule on the graphite matrix DFT calculations were performed. In the case of a high concentration, $C_8Br_2$, the molecule was found to be oriented parallel to the graphene planes with the bromine atoms centred on top of the hexagons (figure 1(a)) and with a calculated equilibrium interlayer distance of 0.64 nm. Both findings are in reasonable agreement with experimental data [11–13]. The calculated cohesive energy per carbon atom is 18 meV $C^{-1}$, which is smaller, but not significantly smaller than for graphite (35 meV $C^{-1}$) [13, 14]. At intermediate concentration, $C_{18}Br_2$, the state with $Br_2$ oriented perpendicular to the graphene planes and the bromine atoms situated on top of a carbon atom was found to be the most stable (figure 1(b)). The theoretical interlayer distance in this situation is 0.77 nm (experimental value 0.88 nm [12]) and the cohesive energy as small as 6 meV $C^{-1}$. For the lowest concentration simulated, $C_{32}Br_2$, the optimal structure was found to be similar to the high concentration case (figure 1(c)) and the interlayer distance 0.62 nm for the whole superlattice without buckling of the graphene sheets. The interlayer binding energy was found to be 3 meV $C^{-1}$, which is almost ten times smaller than for graphite. It is interesting to note that the interlayer binding energy is related to the charge transfer from Br2 to the graphene layers which were determined to be 0.08e, 0.04e and 0.01–0.02e for high, medium and low bromine concentrations, respectively. All these results indicate that a large increase in delamination of the graphite during sonication can be expected at low concentrations of intercalated bromine. Furthermore, at low concentration the cohesive energy of the bromine molecule to the graphene was found to be about 400 meV/Br$_2$. This is roughly four times smaller than for the –OH groups in graphite oxide [6], indicating that Br$_2$ should be less difficult to remove from graphene than –OH groups.

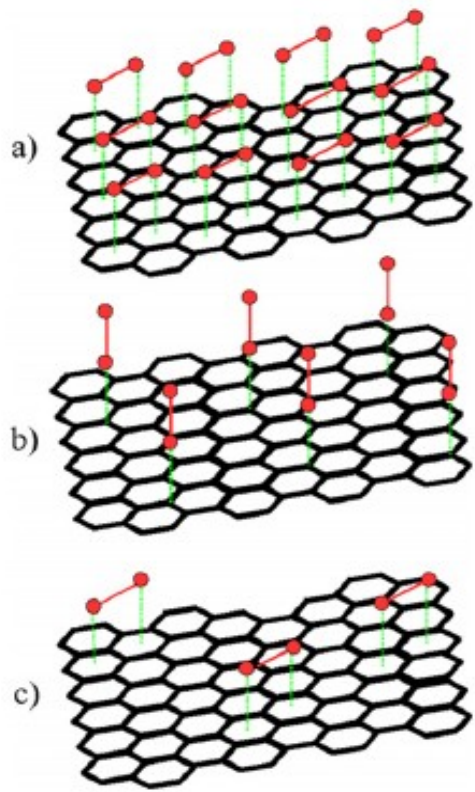

**Figure 1.** Schematic image of the Br$_2$ molecule orientation at different concentrations, (a) C$_8$Br$_2$ (b) C$_{16}$Br$_2$ and (c) C$_{32}$Br$_2$.

Graphite intercalation compounds can be prepared by several different methods [10] and in this study a liquid route was chosen. The Raman spectra of graphite before and after treatment with aq Br$_2$ are shown in figure 2(a). After intercalation, features appear in the 200–1200 nm region originating from the Br2 molecules [15, 16]. The strongest peak is found at 240 cm$^{-1}$ ($\omega_0$) and several harmonics of this line can also be seen. The results from Raman analysis were confirmed by x-ray photoelectron spectroscopy XPS and the concentration of bromine after 48 h of exposure to bromine water was estimated to be approximately 1 at%.

## 2.2. Sonication

Ultrasonic treatment of graphite in water results in small flakes visible to the naked eye floating on the water surface (figure 2(b)). The chemical effects of sonication originate primarily from acoustic cavitation, i.e. the formation, growth and implosive collapse of bubbles in a liquid. This process produces intense local heating, high pressures, enormous heating and cooling rates, and liquid jet streams [17, 18]. Intercalating $Br_2$ into the graphite results in increased flake formation during sonication (figure 2(b)). In accordance with the calculations, this can be explained by a decrease in the cohesive forces between the layers in the graphite due to the presence of the $Br_2$ molecule.

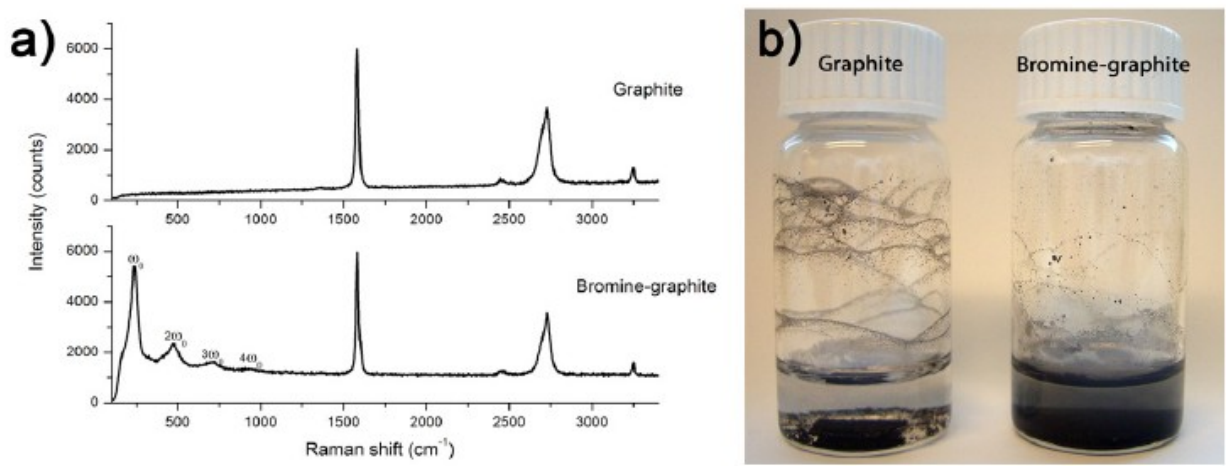

**Figure 2.** (a) Raman spectra of graphite (upper) and bromine–graphite (lower), (b) Photograph of graphite (left vial) and bromine–graphite (right vial) sonicated for 10 min in deionized water.

The flakes produced by the ultrasonic treatment were deposited onto substrates by a simple dipping technique. Figures 3(a) and (b) show scanning electron microscopy (SEM) micrographs of samples prepared using graphite and bromine–graphite, respectively. Because the flakes are hydrophobic and accumulate at the water surface, conclusions about differences in the absolute amount of flakes formed during sonication cannot be drawn. However, SEM analysis (figure 3) indicates a difference in the thickness distribution of the flakes formed using bromine–graphite as compared with graphite. Intercalation of bromine increases not only the amount of material delaminated from the graphite (figure 2(b)), but also the number of thinner flakes formed during sonication (figures 3(a) and (b)). The SEM analysis (figure 3) also indicates a large size distribution of the resulting suspensions with flakes up to about 30 μm, corresponding approximately to the grain size of the starting material. The yield of this process has not yet been determined due to the problems to either collect all flakes or to find a reliable method to measure mass change of the remaining graphite without interference of the intercalated bromine and solvent. However, as expected, a longer sonication time gives a higher yield but this may also damage the material. Preliminary results (not shown here) suggest that sonication times of more than 45 min at 100 W starts to damage the flakes in the suspensions. Further studies of this effect have to be carried out in the future.

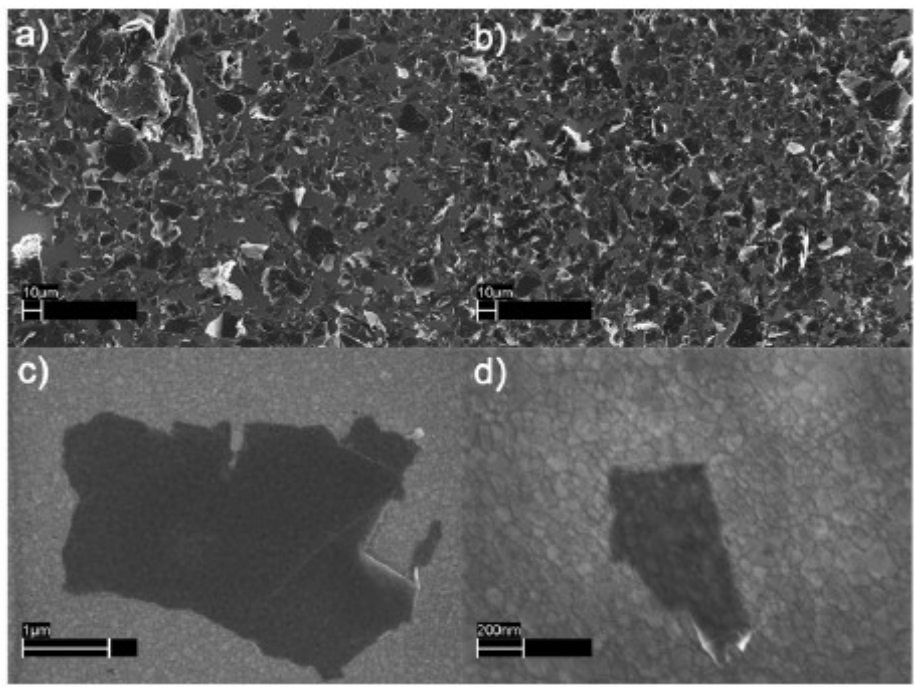

**Figure 3.** Overview SEM-image of samples deposited on silicon from (a) suspensions prepared form graphite and (b) suspensions prepared from bromine–graphite. SEM-images (1.5 kV) of (c) a large flake and (d) a small flake deposited on platinized silicon (note that the structure of the underlying substrate is clearly visible through the flake).

The suspensions were also characterized by transmission electron microscopy (TEM) by collecting the flakes on grids with a carbon support film. Many flakes are folded, making it possible to count the number of layers in high resolution images (figures 4(a) and (b)). This information together with intensity measurements was also used to create thickness maps of entire samples (figures 4(c) and (d)). As indicated by the SEM analysis, a wide distribution was found in both size and thickness but a significant number of sheets <5 layers were observed. These results were also supported by Raman analysis (figure 4(e) [19, 20]). The chemical composition of the flakes in the suspension after intercalation and sonication has not yet been determined but it is likely that, for example, the edges of the graphite planes are terminated by e.g. Hydrogen and/or hydroxyl groups.

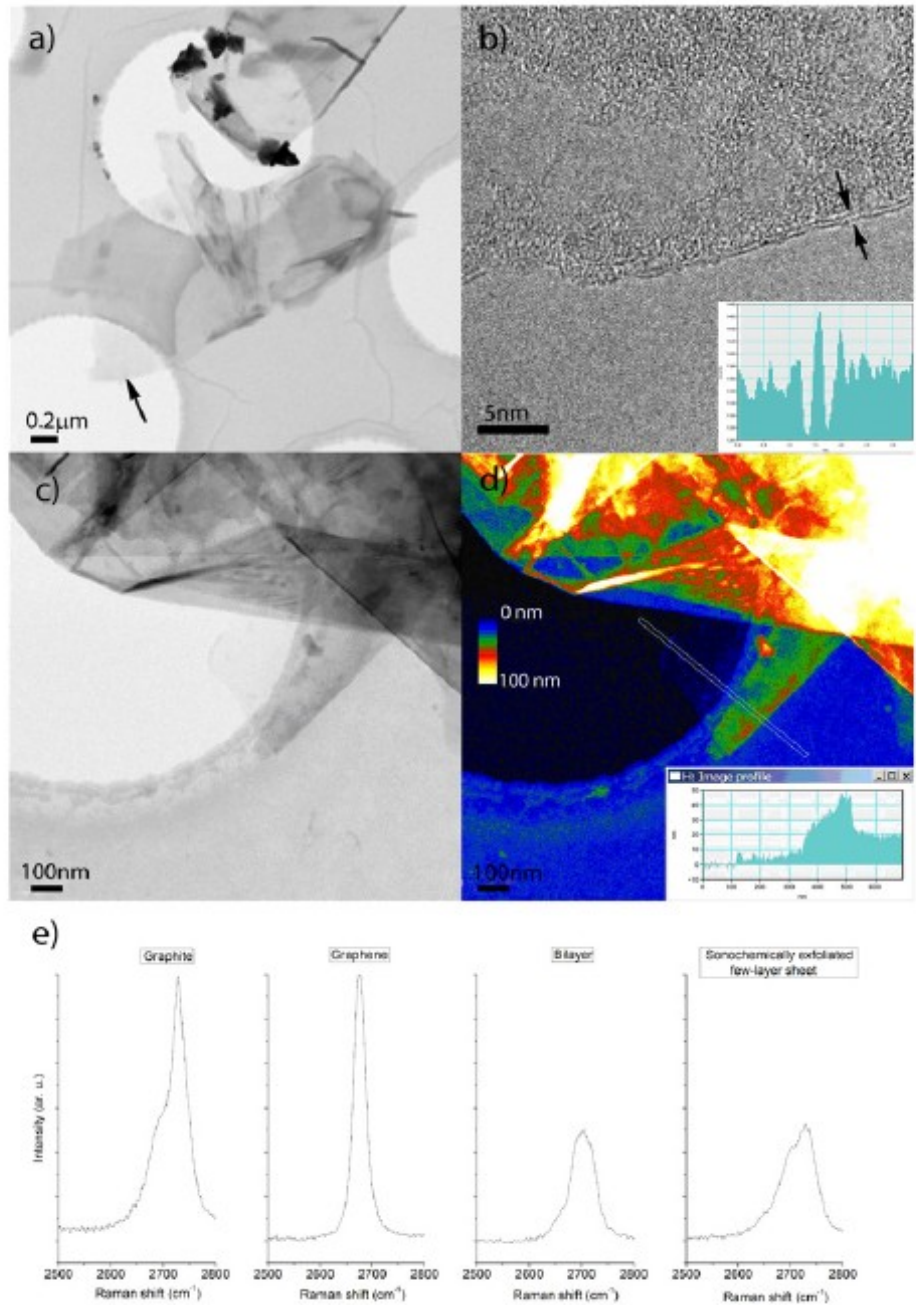

**Figure 4.** (a) An overview of TEM-images (300 kV) of two flakes. (b) High resolution image of the folded edge indicated by the arrow in image (a). The flake is here only 2 or 3 layers thick, as indicated by the intensity profile in the inset. (c) An overview TEM-image (300 kV) of a folded flake and (d) a thickness map constructed from the measured intensities in overview image. The insets show a thickness profile of the marked region in (d). (e) A comparison of the 2D peak of the Raman spectra of graphite foil, reference graphene sample, reference bilayer sample and a sonochemically exfoliated flake.

## 3. Conclusions

To summarize, sonication in combination with $Br_2$ intercalation is a promising route for synthesis of graphene. Calculations show that introducing $Br_2$ can result in an almost ten times lower interlayer binding energy than for graphite, and suspensions with a significant content of few-layer flakes with sizes up to 30 μm have been achieved. Future work will focus on optimizing intercalation and sonication conditions as well as development of separation procedures to achieve dispersions with a narrow thickness and size distribution.

## 4. Experimental section

All chemicals used were from commercial sources and were used as received. Graphite foil (99.8%, metals basis) with a thickness of 0.5 mm was used and cut into 1 cm² pieces before use. Saturated aq $Br_2$ was prepared by mixing bromine with deionized water and the solution was stored in the dark to prevent light-induced addition reactions.

### 4.1. Intercalation

Bromine–graphite was prepared by immersing pieces of graphite (~5 mg) in saturated bromine water (4 ml) for 48 h in a closed vial. Reference samples were prepared using only deionized water. After exposure to the solutions the graphite was allowed to dry in air at RT for ~1 h.

### 4.2. Sonication

A piece of graphite or bromine–graphite was placed in a vial with deionized water (4 ml) and sonicated for 10 min using an ultrasonic bath (45 kHz, 100 W). The remaining piece of graphite/bromine–graphite was removed from the suspension.

### 4.3. Characterization

The graphite and bromine–graphite were characterized by Raman spectroscopy and XPS, utilizing a Renishaw micro-Raman system 2000 and an excitation wavelength of 514 nm, and a PHI Quantum 2000 ESCA respectively. The exfoliated material was also analysed by Raman spectroscopy as well as by a Zeiss LEO 1550 SEM and a FEI Tecnai F30 TEM.

SEM and Raman samples were prepared by dipping silicon or platinized silicon substrates into the suspensions. The substrates were degreased by 5 min sonication, first in trichloroethylene and second in acetone followed by rinsing with ethanol and drying in a flow of $N_2$. After deposition, the samples were rinsed in deionized water and dried in air. In the Raman spectroscopy study, reference samples (Graphene Industries) with a known number of layers were also analysed. TEM samples were prepared by dipping copper grids with a carbon support into the suspensions, rinsing with water and drying in air. The number of graphene layers in a flake can be counted by taking high resolution images of folded regions of the flake. Thickness maps of the samples were also acquired by measuring the intensity of the scattered electron beam. The scattered intensity as a function of thickness is given in linear approximation in equation (1) [21]. A thickness map for the whole sample can be constructed utilizing the difference in intensity recorded from different parts of the sample, and by comparing it with the intensity of the unscattered beam.

$$I(t) = I(0) \cdot 1 - t/\lambda, \qquad (1)$$

where $I(t)$ is the electron counts after passing through a region of thickness t, $I(0)$ is the electron counts in vacuum and $\lambda$ is the characteristic scattering length, which can be estimated by applying equation (1) to areas of known thickness. We obtain a value for $\lambda \sim 225$ nm.

**4.4. Computational details**

To model graphite intercalated compounds, $C_nBr_2$ supercells were used with n = 32, 28 and 8. The interlayer coupling energy was defined as in equation (2),

$$E_{Inter} = (E_{GIC} - E_{CnBr2})/2n, \qquad (2)$$

where $E_{GIC}$ is the total energy of the graphite intercalated compound and $E_{CnBr2}$ is the total energy of single-layer graphene per elementary cell with n carbon atoms and one $Br_2$ on top. The cohesive energy of $Br_2$ on graphene was defined as in equation (3),

$$E_{Bind} = E_{CnBr2} - E_{Cn} - E_{Br2} \qquad (3)$$

where $E_{Cn}$ is the total energy of the supercell of pure graphene and $E_{Br2}$ is the total energy of the bromine molecule. The calculations have been performed using the pseudopotential code SIESTA [22, 23] within the local density approximation (LDA) [24], which is known to be adequate for weakly bonded layered systems [13]. Technical details are close to those used for graphite oxide in [6]. Orientations of bromine molecule parallel and perpendicular to the layers have been considered, as well as positions of bromine atoms on top of carbon atoms and of centres of the hexagons.

**Acknowledgment**

This work was supported by the Swedish Research Council as well as the KOF priority project at Uppsala University